\documentclass[prb,aps,twocolumn,amsmath,amssymb,floatfix,superscriptaddress]{revtex4}
\usepackage[dvips]{graphics}
\usepackage{color}
\definecolor{dred}{rgb}{0,0,0.6}

\begin{document}

\title{Spin half-adder}

\author{Moumita Patra}

\email{moumita.patra19@gmail.com}

\affiliation{Department of Physics, Indian Institute of Technology Bombay,\\
Mumbai, Maharashtra 400076, India}

\begin{abstract}

A new proposal is given to design a spin half-adder in a nano-junction. It is well known that
at finite voltage a net circulating current (known as circular current) appears within a mesoscopic
ring under asymmetric ring-to-electrode interface configuration. This circular current induces
a finite magnetic field at the center of the ring. We utilize this phenomenon to construct a spin
half adder. The circular current induced magnetic field is used to regulate the alignments of
local free spins, by their orientations we specify the output states of the `sum' and `carry'.
All the outputs are spin based, therefore the results get atomically stored in the system. We
also illustrate the experimental possibilities of our proposed model. 

\end{abstract}

\maketitle

\section{Introduction}
 
The ultimate goal of modern technology is to make atomic scale devices. The
continuous shrinking in the size of the channel length of a transistor has driven the industry
from the first four-function calculators to the modern laptops~\cite{datta}. The functionality
of these atomic scale devices are based on the quantum nature of the electrons. But the movement
of charge within an information processing device always associates dissipation which makes the
device energy in-efficient~\cite{SB}. Replacement of ``electrons" by ``spin" is found to be
a most suitable way to resolve these problems. In 1990~\cite{SFET}, Datta and Das came out with
a proposal of spin-field-effect-transistor (SFET), where they used the spin degree of freedom of channel
electrons instead of the charge. Starting from the idea of SFET, till now various proposals have been
reported using spin degrees of freedom, such as spin injection into semiconductors from ferromagnetic
metals~\cite{spin1,spin2,spin3,spin4,spin5}, the development of diluted ferromagnetic
semiconductors~\cite{spin6,spin7}, etc. These devices have several advantages like,
low power consumption, speedy processing, etc. compared to the commonly used semi-conducting
devices~\cite{ref6,ref7,ref8,ref9}. Apart from these advantages, the most important factor in the
context of computation is that these devices are non-volatile in nature. Therefore, unlike charge
based microprocessor, the spintronic devices can store the output itself and we do not need any extra
memory device. For example, using magneto-resistive elements, AND, OR, NAND, and NOR gates have
been constructed with non-volatile output~\cite{ref4}. Dery {\it et. al} have designed logic gate
that consisted of a semi-conductor structure with multiple magnetic contacts~\cite{lg1}. In a recent
work Datta {\it et al.}~\cite{LGS5} have proposed all spin logic devices along with storage mechanism.
Spin-orbit interaction has been used to perform a universal logic operation utilizing minimum
possible devices~\cite{lg2}. In another work A. A. Khajetoorains {\it et al.}~\cite{lg3} have combined
bottom-up atomic fabrication with spin-resolved scanning tunneling microscopy to construct and readout
atomic-scale model systems performing logic operations.

It is always important to design spin based combinational digital circuit at atomic scale. In this article
we propose a spin half-adder using circular current induced magnetic field
\begin{figure}
{\centering\resizebox*{8cm}{6cm}{\includegraphics{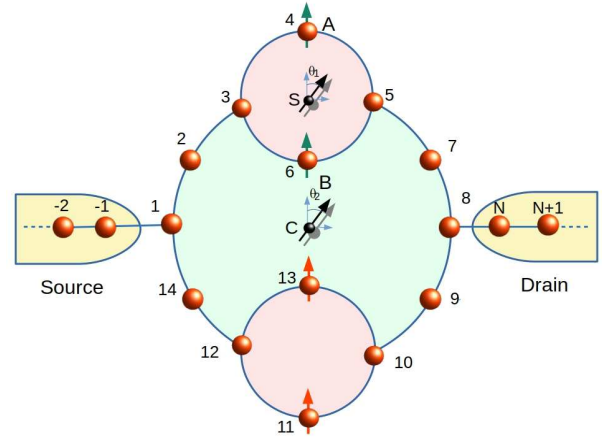}}\par}
\caption{(Color Online). Model of the half-adder where a quantum channel containing multiple
loops is connected to the electrodes. The spin orientations
of sites 4 and 6 is taken as inputs, whereas the sum and carry are specified by the alignment
of free spin namely S and C, respectively.}
\label{fig1}
\end{figure}
in a comfirmational interface. An usual half-adder consists of an AND and XOR gates which are
independently composed of various transistors, resistors, capacitors, etc. Whereas our model has a
strikingly simple design consisting of couple of loops, where the orientations of spins denotes the
low and high states of the inputs and outputs. 

Under finite bias condition, a net current~\cite{ref10,ref11,ref12,ref13,ref14}
appears within the ring, along with the transport current (or drain current). This current
is known as circular current $I$. Circular current is analogous to the persistent current
in a Aharonov-Bhom ring, where the driving force is magnetic field. The circular current
produces a net local magnetic field $B$ at the ring center. In some cases
the magnitude of $B$ reaches $\sim$Millitesla (mT) even up to the order of $\sim$T. Such high
local magnetic field can be used to manipulate the alignment of a local spin embedded at
the center of the ring or at any point on the axis (say, $Z$-axis) passing through the
center of the ring~\cite{ref11,ref13,ref15}. Using the bias induced
circular current, here we design a half-adder where all the inputs and outputs are spin based.
The system is composed of a nano-channel sandwiched between electrodes, namely source and
drain (as shown in Fig.~\ref{fig1}). The electrodes are semi-infinite and non-magnetic.
The channel consists of multiple loops. The spin orientations (viz up and down) of sites 4 and
6 are taken as inputs for the entire operation. The outputs of the half-adder i.e., `sum'
and `carry are specified by the spin orientations of two free spins S and C, respectively
which are attached at the center of the top loop containing atomic sites 3-4-5-6, and at
the center of the whole system, respectively. The sum and carry are not
orientated at the $Z$-direction and make an angle $\theta_C$ (say, $\theta_C = 30^{\circ}$).
This can be done by the application of a constant external magnetic field. We use the
circular current induced magnetic field to tune these free spins. Under finite bias
condition the circular current so as the magnetic field vanish when the loops are symmetric,
\begin{figure}[ht]
{\centering\resizebox*{5cm}{3cm}{\includegraphics{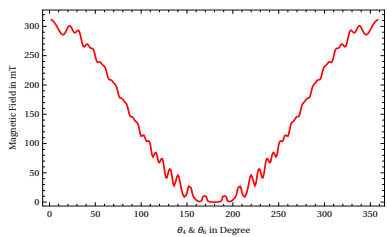}}\par}
\caption{(Color online) Circular current induced magnetic field at the center
of the device as a function of $\theta_4$ and $\theta_6$ for $\theta_{11} = \theta_{13} = 180^{\circ}$
(The bias voltage $V = 0.75 V$ and the temperature $T = 100\,$K).}
\label{mag}
\end{figure}
and they reappear for asymmetric loop geometry. This is the key idea behind the logical
operations. When a large magnetic field is produced in the loop, the
corresponding free spin changes its orientation towards $Z$-direction. Whereas when
the circular current (and the associated magnetic field) vanishes, the free spin again moves
back by an angle $\theta_C$ due to the applied external constant magnetic field. In the system
the asymmetry is introduced by the different spin
orientations of sites 4 and 6.
\begin{figure*}
{\centering\resizebox*{17cm}{8cm}{\includegraphics{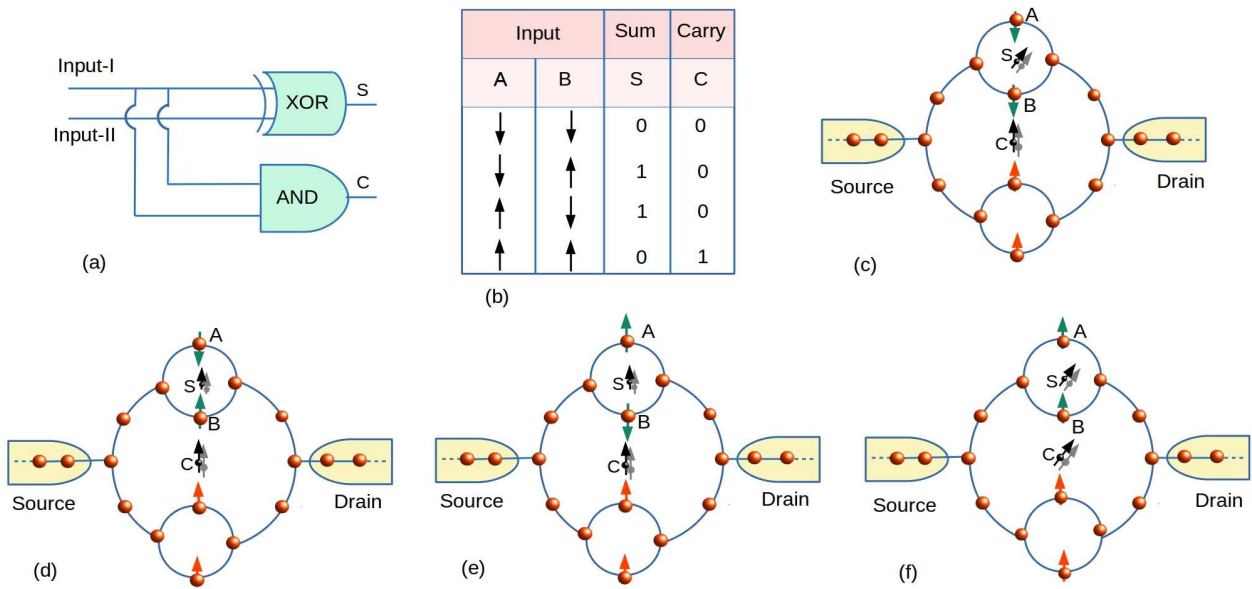}}\par}
\caption{(Color Online). The operational principle of the spin half-adder. (a) Sketch of
standard electronics half-adder consist of an XOR and AND gate. (b) Truth table for spin
half-adder. (c) - (f) The  representation of all the spin based half-adder operations
for four input conditions i.e., (0,0), (0,1), (1,0), and (1,1), respectively. Here the
input states are specified by the spin orientations of A and B (shown in green
color). Whereas the sum and carry are specified by free spins S and C (shown in
black color), respectively.}
\label{fig2}
\end{figure*}
As the Boolean operation is based on the manipulation of local spin by the
means of circular current, this device is expected to has negligible power-consumption and delay
time~\cite{mp}, and very high endurance ($> 10^{15}$ cycle) which exceeds the requirements of various
memory use cases, including high-performance applications such as CPU level-2 and level-3
caches~\cite{en}, as we find in the spin-transfer-torque random-access memory (where local spin
is regulated by spin polarized current).

An efficient readout of electronic spins denoting the outputs sum and carry, is
required for the experimental execution. The traditional magnetic resonance techniques rely on large
ensembles of nuclear spins. Though the ultimate goal is the readout of the single spin states and there
are lots of proposals available in this direction at the present time. For example,
spin readout of nitrogen-vacancy (NV) centers in diamond is achieved by the conversion of the
electronic spin state of the nitrogen-vacancy to a charge-state distribution, followed by
single-shot readout of the charge state~\cite{sr1}. A versatile single spin meter is designed which
is consisting of a quantum dot in a magnetic field under microwave irradiation combined with a charge
counter~\cite{sr2}. Several other proposals are also available based on the physical system in which
the spin is housed~\cite{sr3} or requiring certain special features, such as optical activity~\cite{sr4},
nuclear spins~\cite{sr5} or a large detector-system interaction~\cite{sr6}, etc. These proposals make us
confident about the spin state readability of our system.

There are several characteristic features which substantiate the
robustness of the half adder.

\begin{enumerate}

\item Though in this paper we consider a simplified model of 14 atomic sites,
but the results are equally true for any same kind of system having more atomic sites. But
in each loop the number of atoms should be even such that they can be symmetrically
connected to the other part of the circuit.
	
\item So far in the literature, to construct spin logic gates, the spins are used
either as input or output variables. So spin-to-charge converter
is required for every operation,  which causes the loss of efficiency. As in
our set up  all the operations are described by only spin states, no
spin-to-charge conversion is required.

\item The spin based logic devices can store the information which is very important for
non-volatile computations in computer. On the other hand the conventional charge based
computers are volatile.

\item The spin injection efficiency and the material of the channel (semi-conductor or
metal) are two extremely important aspects for any experimental application~\cite{ref14a}. 
Metal has high spin injection efficiency~\cite{LGS5}, but the spin coherence length is
smaller for a metallic channel. On the hand, semi-conductor has high coherence length, but
inadequate spin injection efficiency. In this paper, the proposed model has a metallic channel
with smaller length (only 14 atomic sites), so that it has high spin injection efficiency,
whereas the issue of coherence is solved because of its smaller system size.
		
\item The proposed model can be reprogrammed to have various other logical
operations like NAND, NOT, OR, etc.

\item As the spin states are solely described by the up and down spins so, an
efficient mechanism to rotate local spins is required. We describe an experimental setup in
section.~V, where we use the bias induced magnetic field~\cite{ref15,cite22} to regulate
the input states.

\item The results are valid at non-zero temperature, which is very crucial for
practical applications.
		
\end{enumerate}

We arrange the paper in following manner. The theoretical method is discussed in Sec~II.
In Sec~III we present all the results.
In Sec~IV, various other logical operations are demonstrated in the same setup.
An experimental proposal is demonstrated in Sec~V and finally, we give a over view in
Sec~VI.

\section{Theoretical prescription}

To calculate the circular current in a nano junction, we use the wave-guide theory.
We start by writing the tight binding Hamiltonian of the model as shown in Fig.~\ref{fig1}.
The system consists of quantum channel connected to the electrodes. Therefore the Hamiltonian
for the entire system becomes
\begin{equation}
H= H_C + H_S + H_D + H_T
\label{eq1}
\end{equation}
Here the $H_C$, $H_S$ and $H_D$ the Hamiltonians for the channel (C),
source (S) and drain and they read as:
\begin{equation}
H_{\alpha}=\sum \epsilon_{\alpha,\sigma}c_{n,\sigma}^{\dagger}c_{n,\sigma} +
\sum\left(c_{n+1,\sigma}^{\dagger} t_{\alpha,\sigma} c_{n,\sigma} + h.c. \right)
\label{eq2}
\end{equation}
\noindent where $\alpha=C, S, D$. For the electrodes, the on-site energy and
atom-to-atom coupling become : $\epsilon_{\alpha,\sigma}=\epsilon_0$ and
$t_{\alpha,\sigma}=t_0$, respectively whereas for the channel they are
$\epsilon$ and $t$, respectively. Sites 4, 6, 11,
and 13 are magnetic. The onsite potential for these sites are:
$\epsilon - h_i.\sigma$ ($h_i = \sim |h_i|$ represents the spin-flip scattering
and $\sigma$ is Pauli matrices). The last term $H_T$ of Eq.~\ref{eq1}
is the tunneling Hamiltonian. 

In the channel, the atomic sites 4, 6, 11, and 13 are magnetic, we need to calculate
all spin dependent components of circular current $I_C$. In one of our recent
works~\cite{ref14} we have put forward the methodology to calculate the spin components of
circular current. Here we follow the same
prescription.The detailed calculation of the bond current density $J_{i,i+1}$ between
the site $i$ and $i+1$ is described in Appendix~\ref{aa}. The current at bias voltage $V$ can
be written as~\cite{ref16,ref17}
\begin{equation}
I_{i,i+1}(V) = \int\limits_{-\infty}^{\infty}J_{i,i+1}(E)[f_S(E) - f_D(E)]\, dE
\label{eq6}
\end{equation}
Where, $f_{S(D)}=\Big[1 + e^{\frac{E-\mu_{S(D)}}{k_B T}}\Big]^{-1}$ is
the Fermi function ($k_B$ is the Boltzmann constant and $T$ is the temperature)
corresponding to the source and drain and $\mu_{S(D)}$ is the corresponding
chemical potential.

In the present system we need to calculate the circular current for two loops: one is for the
top loop containing atomic sites 3, 4, 5, and 6, and other for the center loop containing
atomic sites 1-2-3-6-5-7-8-9-10-13-12-14. After calculating the corresponding bond currents
we calculate net circular current of these two loops as:
\begin{equation}
I_1 = \frac{1}{4}\left(I_{3,4} + I_{4,5} + I_{5,6} + I_{6,3}\right)
\label{eq7a}
\end{equation}
and
\begin{eqnarray}
I_2 & = & \frac{1}{12}(I_{1,2} + I_{2,3} + I_{3,6} + I_{6,5} + I_{5,7} + I_{7,8}
+ I_{8,9} \nonumber \\
& & + I_{9,10} + I_{10,13} + I_{13,12} + I_{12,14} + I_{14,1})
\label{eq7b}
\end{eqnarray}
respectively. If the current goes in anti-clockwise direction then we consider it positive,
and vise-versa. 

Net local magnetic fields are established as the circular currents flow with in the rings.
Using the Biot-Savart's law we can calculate the magnetic fields as
\begin{equation}
\vec{B_n}(\vec{r_n}) = \sum\limits_{\langle i,j \rangle} \left(\frac{\mu_0}{4\pi}
\right)
\int I_{i,j}\frac{d\vec{r^{\prime}} \times(\vec{r_n}-\vec{r^{\prime}_n})}
{|(\vec{r_n}-\vec{r^{\prime}_n})|^3}
\label{eq8}
\end{equation}
$n=1,2$. $\mu_0$ is the magnetic constant. 

Now we consider a free spin is embedded at the ring center as shown
in Fig.~\ref{fig1}. The spin is initially misaligned with $Z$
direction. With the appearance of $I_C$ and associated $B$, the spin
tries to align itself along $Z$ direction. We calculate the spin
angle of rotation $\theta_C$~\cite{ref13,cite12,cite22} by the magnetic field $B$
for a time $\tau$ as
\begin{equation}
\theta_i=g \mu_B B \tau /2\hbar
\label{eq9}
\end{equation}
$i=1,2$. $g(\approx1) \rightarrow$ Lande $g$-factor; $\mu_B \rightarrow$ Bohr magneton.

\section{Numerical Results and Discussion}

As the functionality of the half-adder depends on the appearance of
circular induced magnetic field in asymmetric situation, we want to examine the dependence of the
induced magnetic field on the system-asymmetry. In Fig.~\ref{mag}, we plot the magnetic field
produced at the center
of the device with the angle of rotation of the inputs-A and B. We consider spin-11 and 13
to be down. All the $h_i$'s are $0.25\,$eV. Here we find that a large magnetic field is produced
when the system has the most asymmetry and it smoothly varies towards zero as the
orientations of A and B become similar to spins-11 and 13, and this is the key factor of our proposal.
For the execution of logic operation the appearance of zero circular current is a prime requirement.
This demands an ideal symmetric condition, which seems to be unrealistic in
real situation. But we can see in Fig.~\ref{mag} that the produced magnetic field is insufficient to
rotate the spin along $Z$ for a considerable range around the symmetry point
($\theta_4 = \theta_6 = 180^{\circ}$). Therefore we can argue that, even if there is little
asymmetries due to manufacturing imperfection and other factors, but this proposal will
still be equally valid.

Now We explain the half-adder operation.
\begin{figure}[ht]
{\centering\resizebox*{8cm}{5cm}{\includegraphics{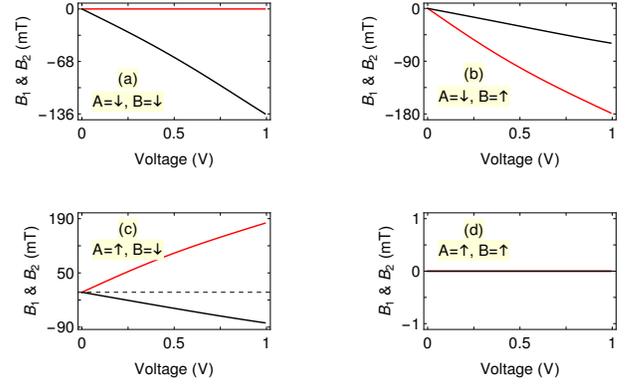}}\par}
\caption{(Color online) (a)-(d) Produced magnetic fields $B_1$ (red curve) and
$B_2$ (black curve) associated with the sum and carry, respectively for the four input conditions.
Here we set $T = 250\,$K.}
\label{fig3}
\end{figure}
In Fig.~\ref{fig2} we present the circuit diagram and the truth table for spin half adder.
The input states are specified by the spin orientations of
spin-A and B as $\downarrow : \rightarrow 0$ and $\uparrow : \rightarrow 1$.
Whereas the output conditions are specified by the spin alignment of free
spin S (represents sum) and C (i.e., carry). The mechanisms
of sum and carry are as follows:

\vskip 0.1cm
\noindent
{\bf \underline{Sum}:} We assume that the free spins initially are not aligned
along $Z$-direction. The output for the sum is defined as: if the free spin S is in its
initial position, then the output is 0, and if it is aligned along $Z$, it
represents 1. The alignment of the individual free spin depends
on the appearance of current induced magnetic field in
each loop. So  there is no magnetic field when both A and B are
parallel, i.e., when both are either up or down (see Fig.~\ref{fig2}(c) and
\begin{table}[ht]
\caption{Truth tables for different parallel logical operations.}
$~$
\vskip -0.25cm
\fontsize{7}{11}
\begin{tabular}{|c|c|c|c|}
\hline
\multicolumn{2}{|c|}{Input} & {Sum ($|B_1|$ in mT)} & {Carry ($|B_2|$ in mT)}\\
\hline
A & B & S & C \\
 \hline
$\downarrow$ & $\downarrow$ & 0 &  62.7\\
$\downarrow$ & $\uparrow$ & 101 & 30.3 \\
$\uparrow$ & $\downarrow$ & 101 &  41.5\\
$\uparrow$ & $\uparrow$ & 0 & 0 \\
 \hline
\end{tabular}
\label{tab1}
\end{table}
(d)). In these cases S remains in its initial position such
that logical output becomes 0. When they are anti-parallel (shown in
Fig.~\ref{fig2}(e) and (f)) the corresponding loop becomes asymmetry and a
net magnetic field is produced along $Z$ direction and the free spin S follows
the field. These situations imply 1. So, this part of the circuit behaves an
XOR gate which is the sum of the half adder.

\vskip 0.1cm
\noindent
{\bf \underline{Carry}:} To generate carry we consider the
full system, and place a free spin C at the center of the whole system.
The output for the carry is assumed to be 0 when C is aligned along $Z$,
otherwise it is 1. As the lower loop contains two up spins,
the whole system becomes symmetric only when
both the inputs (i.e., A and B in upper loop) are up. So no magnetic field is
developed at the center of the circuit and C remains in its initial position
(Fig.~\ref{fig2}(f)). This situation implies 1. For all three input conditions
the system is asymmetric and a net magnetic field produced at the center which
turns the free spin C along $Z$ direction (as shown in Fig.~\ref{fig2}(c) - (e))
hence the output becomes $0$ in these cases. So AND behavior is accomplished at C
and hence the construction of half adder is accomplished. 

In Fig.~\ref{fig3} we plot the produced magnetic fields $B_1$ and $B_2$ associated
with the operation sum and carry, respectively as a function of voltage for
four different input conditions (i.e., when inputs A and B are: $(\downarrow, \downarrow)$,
$(\downarrow, \uparrow)$, $(\uparrow, \downarrow)$, and $(\uparrow, \uparrow)$ and shown in
Fig.~\ref{fig3}(a)-(d), respectively). Here we set, all site energies to zero, hopping
integral in contacting leads at $1\,$eV, and all the $t_i$s' in the ring at $0.5\,$eV and
\begin{figure}[ht]
{\centering\resizebox*{8cm}{3cm}{\includegraphics{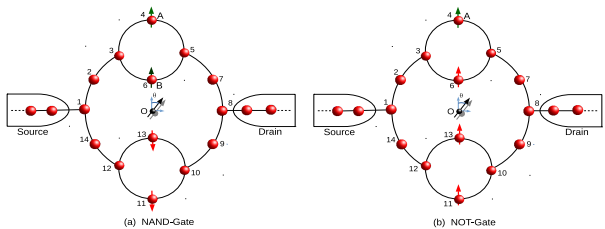}}\par}
\caption{(Color online) Layouts of different logic gates.}
\label{lg}
\end{figure}
the ring-to-lead couplings at $0.5\,$eV. The magnitude of the net magnetization at sites
4, 6, 11, and 13 are $0.5\,$eV. The calculations are done considering $250\,$K temperature. The
average atomic distance $a$ is considered to be $1~\AA$. The red curve represents the magnetic
field corresponding to sum (i.e. $B_1$) whereas black one represents $B_2$ which is the
magnetic field associated to carry. Let the free spins corresponding to S and C  are
initially set at $30^{\circ}$ . When the loops become asymmetric, the circular currents
hence net magnetic fields will produce in each loop that will turn the free spins S and
C towards $Z$ direction. Considering the desired operation time as $\tau=5\,$ns~\cite{cite22},
we can
calculate the desired magnetic field to align the spin along $Z$ is $\sim 2.4\,$mT (as
follows from Eq.~\ref{eq9}). So we need at least $\sim 2.4\,$mT to execute all the logic
operations. For all the four cases of Fig.~\ref{fig3} we find large enough magnetic fields
are produced which are more than sufficient to turn S and C in appropriate cases. On the
other hand for the proper cases the magnetic fields are exactly zero, which will leave
the S and C in its initial positions. As the results remain valid for large ranges of voltage
and temperature, we can expect that the proposed model might be implemented in the
laboratory. 

The quantitative representation of half adder is shown in Table~\ref{tab1}. Here all the
parameters are chosen to be the same as Fig.~\ref{fig3} and magnetic fields are evaluated at
bias voltage $0.5\,$V.

\section{Reprogrammable spin logic gate}

In this paper our main motivation is to construct spin half-adder, though
other logical operations can be accomplished by reprogramming the same system. For
example in Fig.~\ref{lg}, we have shown the sketches for NAND and NOT gates. The
spins shown in green color represent the inputs and O represents the output. As the
logical operations follow the symmetry conditions of the ring, we accordingly
set the spin orientations of the other sites in the system. For NAND gate (Fig.~\ref{lg}(a))
spin-11 and 13 are set to be down. So only for ($\downarrow$, $\downarrow$) input condition,
there is no circular current appears at the center of the system and the output becomes
zero. But for all other three cases the outputs are 1, which is a NAND gate response.
For NOT gate (Fig.~\ref{lg}(b)), there is one input i.e., A, and other spin are considered
to be up. So if A is down, output becomes 1 and vice-versa.

In a similar fashion, we can reprogram this model to have other logical operations
also, which definitely implies the versatility of our proposal.

\section{Experimental setup of half adder}

To have the input conditions 0 and 1, we need to align the
spins in the magnetic ring selectively several prescriptions are available to control single electron spin.
For instance, using radio frequency pulses these spins can be manipulated~\cite{pl22,pl23,pl24},
though in this case relatively larger time scale is required to operate the spins. On the other hand,
\begin{figure}[ht]
{\centering
\resizebox*{4cm}{4cm}{\includegraphics{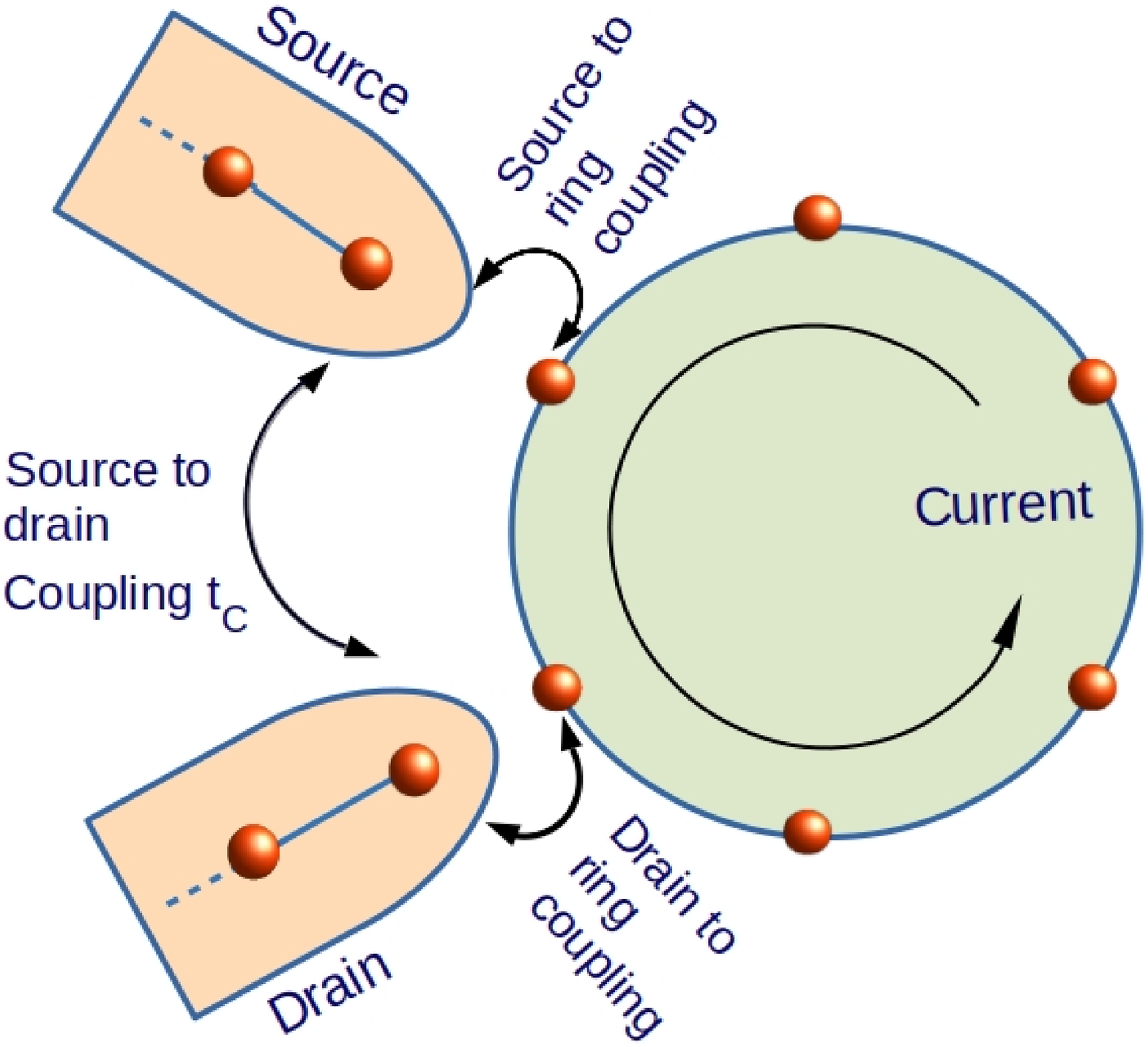}}
\resizebox*{4cm}{2.5cm}{\includegraphics{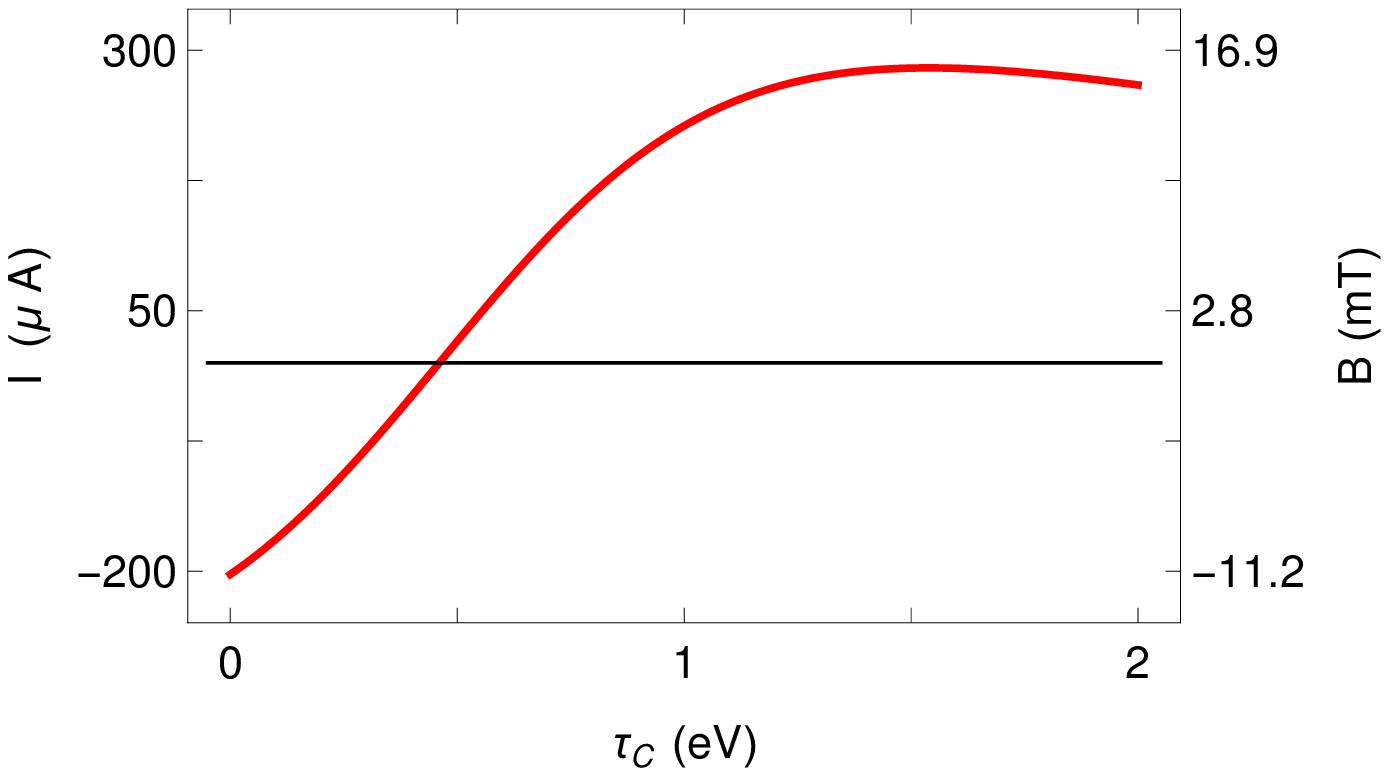}}\par}
\caption{(Color online). Left figure: A nano ring is connected to the electrodes in most
asymmetrically. As there exists a shunting path between the electrodes S and D, electrons
can directly hop between the them. Right figure: Modulation of $I$  and
magnetic field $B$  with the hopping parameter $t_C$. The ring has 20 atomic sites,
its radius is $10$\AA$\,$. $B$ is calculated at a distance $20\,$\AA$\,$ from the
center of the ring to the point of measurement. The applied voltage is
considered to be $V=0.5\,$V. The onsite potential $\epsilon_r$ and hopping integral
$t_r$ for the entire system are considered as: $\epsilon_r=0$ and $t_r=1\,$eV.}
\label{f5}
\end{figure}
the manipulations can be made much faster such as in the picosecond or femtosecond time scale with
the help of optical pulses~\cite{pl25,pl26,pl27}. In another pioneering work Press et al. have shown that
the selective tuning of electron spins are possible within the spins' coherence times by means of
ultrafast laser pulses~\cite{pl28}. With the availability of these various sophisticated prescriptions, we
strongly believe that the alignment of selective spins (i.e., 4 and 6) can be properly adjusted.

Apart from the above mentioned proposals here we present another suitable method for the proper regulations
of the spin specifying input conditions. We make use of the circular current induced magnetic field to
regulate the inputs and for that matter tuning of the magnetic field externally is required.
We take a nano ring with two side attached metallic electrodes (S and D) connected
to the two adjacent sites of the ring as shown in the left figure of Fig.~\ref{f5}.
As the electrodes are connected at the two neighboring sites of the ring,
with a finite probability, the electrons can directly hop from S to D. Let $t_C$ be
related hopping integral between the electrodes. By changing relative positions of the
electrodes we can tune $t_C$, and tuning $t_C$ we can regulate the circular current
induced magnetic field $B$ for a large scale. This proposal has already been discussed
in on of our previous work~\cite{ref21}. For an example, taking an 20-site ring we
show the variation of
the magnetic field with source-to-drain coupling $t_C$ in the right figure
of Fig.~\ref{f5} where we follow the same theoretical prescription for the calculation
of current and magnetic filed as described in Sec.~II. From the result we can conclude
that regulating the tunneling between the side attached electrodes the local magnetic $B$
field can be tuned a large scale which is required for flipping of spin states (i.e., up or
down). Such nano junctions needed to be put on the sites 4 and 6 (as shown in
in the right of Fig.~\ref{f5}, the distance of the ring center to the sites 4 and 6
will be $20\,$\AA$\,$) and in every cases changing the shunting paths between
source-to-drain we can specify the required input conditions.

\section{Closing Remarks}

In conclusion, we have theoretically realized a spin half-adder where all the
inputs are output conditions are completely spin based. The outputs preserve the
memories. The logic operations have been implemented in a nano channel containing multiple
loops. The basic mechanism of replies on the appearance
of circular current and associated magnetic field under a finite bias condition.
We have used the magnetic field to rotate free spins embedded at the center of the loops.
By the proper chooses of their orientations we have specified the 0 and 1 states of `sum' and
`carry' of the half-adder. The proposal is well tested at non-zero temperature and a well
suited experimental setup is discussed for input spins regulations. We have
shown various other logical operation in the same setup utilizing the bias induced circular
current.

In a real situation, many possible sources are there those may destroy
phase and spin memory of electrons, and among them the most common source is electron-phonon
(e-ph), the stray field, and other impurities. Theoretically one can incorporate these effects
by studying the dephasing effecting on the current by introducing phenomenological voltage
probes into the system. The effect of dephasing and impurities on circular current are
different compared to the transport current, which generally decreases with these factors.
On the other hand, circular current may increase in the presence of dephasing
and disorder~\cite{ref14}. Therefore we strongly believe that the results presented here
will be still valid in real experiment.

Though results displayed here are calculated for a specific system, but this proposal will be well
suited for same kind of geometry having any arbitrary number of atomic sites in each loop. The spin
half adder along with other logical operations, discussed here will certainly boost the new
generation computations along with storage mechanism.

\section{Acknowledgements}

The author gratefully acknowledge the fruitful discussions with Prof. Alok Shukla and
Prof. S. K. Maiti. The work has been done with the financial support (post-doctoral
fellowship) from Indian Institute of Technology, Bombay, India.

\appendix

\section{Circular current density}
\label{aa}

The wave guide formalism~\cite{ref14,ref18,ref19,ref20,ref21} involves a set of linear coupled
equations which are obtained from the Schr\"{o}dinger equation $H|\phi\rangle = E|\phi\rangle$
with $|\phi\rangle = \left[\sum A_{n,\sigma} a_{n,\sigma}^{\dagger} + \sum B_{n,\sigma}
b_{n,\sigma}^{\dagger} + \sum C_{i,\sigma} c_{i,\sigma}^{\dagger}\right]|0\rangle$.
The coefficients $A_{n,\sigma}, B_{n,\sigma}$, and $C_{n,\sigma}$
are the wave-amplitude corresponding to the $n$th site of the electrodes (namely, source
and drain), where as $i$ represent the site index of the ring. Let an up spin injected from the
source to the channel as a plane wave with unit amplitude. For our present setup as shown in
Fig.~\ref{fig1}, we have the equations as follows:
{
\allowdisplaybreaks
\begin{widetext}
{\footnotesize
\begin{eqnarray}
\left[\left(\begin{array}{cc}
        E & 0 \\
        0 & E
\end{array}\right) - \left(\begin{array}{cc}
    \epsilon_0 & 0 \\
    0 & \epsilon_0
\end{array}\right)\right]\left(\begin{array}{cc}
        1 + \rho_{\uparrow\uparrow} \\
        \rho_{\uparrow\downarrow}
    \end{array}\right) = \left(\begin{array}{cc}
    t_0 & 0 \\
    0 & t_0
\end{array}\right) \left(\begin{array}{cc}
    e^{-ika} + \rho_{\uparrow\uparrow}e^{ika} \\
        \rho_{\uparrow\downarrow}e^{ika}
    \end{array}\right) + \left(\begin{array}{cc}
   t_{S}  & 0 \\
    0 & t_{S}
\end{array}\right) \left(\begin{array}{cc}
   C_{1,\uparrow\uparrow}  & 0 \\
    0 & C_{1,\uparrow\downarrow}
\end{array}\right)\nonumber \\
\left[\left(\begin{array}{cc}
    E & 0 \\
    0 & E
\end{array}\right) - \left(\begin{array}{cc}
    \epsilon & 0 \\
    0 & \epsilon
\end{array}\right)\right] \left(\begin{array}{cc}
   C_{1,\uparrow\uparrow}  & 0 \\
    0 & C_{1,\uparrow\downarrow}
\end{array}\right) = \left(\begin{array}{cc}
    t_{S} & 0 \\
    0 & t_{S}
\end{array}\right)\left(\begin{array}{cc}
    1 + \rho_{\uparrow\uparrow(S)} \\
    \rho_{\uparrow\downarrow(S)}
    \end{array}\right) + \left(\begin{array}{cc}
    t & 0 \\
    0 & t
\end{array}\right) \left(\begin{array}{cc}
   C_{2,\uparrow\uparrow}  & 0 \\
    0 & C_{2,\uparrow\downarrow}
\end{array}\right) + \left(\begin{array}{cc}
    t & 0 \\
    0 & t
\end{array}\right) \left(\begin{array}{cc}
   C_{14,\uparrow\uparrow}  & 0 \\
    0 & C_{14,\uparrow\downarrow}
\end{array}\right)
\nonumber \\
\left[\left(\begin{array}{cc}
    E & 0 \\
    0 & E
\end{array}\right) - \left(\begin{array}{cc}
    \epsilon & 0 \\
    0 & \epsilon
\end{array}\right)\right] \left(\begin{array}{cc}
   C_{2,\uparrow\uparrow}  & 0 \\
    0 & C_{2,\uparrow\downarrow}
\end{array}\right) = \left(\begin{array}{cc}
    t & 0 \\
    0 & t
\end{array}\right) \left(\begin{array}{cc}
   C_{1,\uparrow\uparrow}  & 0 \\
    0 & C_{1,\uparrow\downarrow}
\end{array}\right) + \left(\begin{array}{cc}
    t & 0 \\
    0 & t
\end{array}\right) \left(\begin{array}{cc}
   C_{3,\uparrow\uparrow}  & 0 \\
    0 & C_{3,\uparrow\downarrow}
\end{array}\right)\nonumber \\
\left[\left(\begin{array}{cc}    
    E & 0 \\
    0 & E
\end{array}\right) - \left(\begin{array}{cc}
    \epsilon & 0 \\
    0 & \epsilon
\end{array}\right)\right] \left(\begin{array}{cc}
   C_{3,\uparrow\uparrow}  & 0 \\
    0 & C_{3,\uparrow\downarrow}
\end{array}\right) = \left(\begin{array}{cc}
    t & 0 \\
    0 & t
\end{array}\right) \left(\begin{array}{cc}
   C_{2,\uparrow\uparrow}  & 0 \\
    0 & C_{2,\uparrow\downarrow}
\end{array}\right) + \left(\begin{array}{cc}
    t & 0 \\
    0 & t
\end{array}\right) \left(\begin{array}{cc}
   C_{4,\uparrow\uparrow}  & 0 \\
    0 & C_{4,\uparrow\downarrow}
\end{array}\right) + \left(\begin{array}{cc}
    t & 0 \\
    0 & t
\end{array}\right) \left(\begin{array}{cc}
   C_{6,\uparrow\uparrow}  & 0 \\
   0 & C_{6,\uparrow\downarrow}
\end{array}\right) \nonumber \\
\left[\left(\begin{array}{cc}
    E & 0 \\
    0 & E
\end{array}\right) - \left(\begin{array}{cc}
    \epsilon + h_4\cos\vartheta_4 & \sin\vartheta_4e^{-i\varphi_4} \\ 
    \sin\vartheta_4e^{i\varphi_4} & \epsilon - h_4\cos\vartheta_4
\end{array}\right)\right] \left(\begin{array}{cc}
   c_{4,\uparrow\uparrow}  & 0 \\
    0 & c_{4,\uparrow\downarrow}
\end{array}\right)
  = \left(\begin{array}{cc}
    t & 0 \\
    0 & t
\end{array}\right) \left(\begin{array}{cc}
    C_{3,\uparrow\uparrow}  & 0 \\
    0 & C_{3,\uparrow\downarrow}
\end{array}\right) +  \left(\begin{array}{cc}
    t & 0 \\
    0 & t
\end{array}\right) \left(\begin{array}{cc}
    C_{5,\uparrow\uparrow}  & 0 \\
    0 & C_{5,\uparrow\downarrow}
\end{array}\right)\nonumber \\
\left[\left(\begin{array}{cc}    
    E & 0 \\
    0 & E
\end{array}\right) - \left(\begin{array}{cc}
    \epsilon & 0 \\
    0 & \epsilon
\end{array}\right)\right] \left(\begin{array}{cc}
    C_{5,\uparrow\uparrow}  & 0 \\
    0 & C_{5,\uparrow\downarrow}
\end{array}\right) = \left(\begin{array}{cc}
    t & 0 \\
    0 & t
\end{array}\right) \left(\begin{array}{cc}
    C_{4,\uparrow\uparrow}  & 0 \\
    0 & C_{4,\uparrow\downarrow}
\end{array}\right) + \left(\begin{array}{cc}
    t & 0 \\
    0 & t
\end{array}\right) \left(\begin{array}{cc}
    C_{6,\uparrow\uparrow}  & 0 \\
    0 & C_{6,\uparrow\downarrow}
\end{array}\right) + \left(\begin{array}{cc}
    t & 0 \\
    0 & t
\end{array}\right) \left(\begin{array}{cc}
    C_{7,\uparrow\uparrow}  & 0 \\
    0 & C_{7,\uparrow\downarrow}
\end{array}\right) \nonumber \\
\left[\left(\begin{array}{cc}
    E & 0 \\
    0 & E
\end{array}\right) - \left(\begin{array}{cc}
    \epsilon + h_6\cos\vartheta_6 & \sin\vartheta_6e^{-i\varphi_6} \\ 
    \sin\vartheta_6e^{i\varphi_6} & \epsilon - h_6\cos\vartheta_6
\end{array}\right)\right] \left(\begin{array}{cc}
    C_{6,\uparrow\uparrow}  & 0 \\
    0 & C_{6,\uparrow\downarrow}
\end{array}\right)
  = \left(\begin{array}{cc}
    t & 0 \\
    0 & t
\end{array}\right) \left(\begin{array}{cc}
    C_{5,\uparrow\uparrow}  & 0 \\
    0 & C_{5,\uparrow\downarrow}
\end{array}\right) +  \left(\begin{array}{cc}
    t & 0 \\
    0 & t
\end{array}\right) \left(\begin{array}{cc}
    C_{3,\uparrow\uparrow}  & 0 \\
    0 & C_{3,\uparrow\downarrow}
\end{array}\right)\nonumber \\
\left[\left(\begin{array}{cc}
    E & 0 \\
    0 & E
\end{array}\right) - \left(\begin{array}{cc}
    \epsilon & 0 \\
    0 & \epsilon
\end{array}\right)\right] \left(\begin{array}{cc}
    C_{7,\uparrow\uparrow}  & 0 \\
    0 & C_{7,\uparrow\downarrow}
\end{array}\right) = \left(\begin{array}{cc}
    t & 0 \\
    0 & t
\end{array}\right) \left(\begin{array}{cc}
    C_{5,\uparrow\uparrow}  & 0 \\
    0 & C_{5,\uparrow\downarrow}
\end{array}\right) + \left(\begin{array}{cc}
    t & 0 \\
    0 & t
\end{array}\right) \left(\begin{array}{cc}
    C_{8,\uparrow\uparrow}  & 0 \\
    0 & C_{8,\uparrow\downarrow}
\end{array}\right)\nonumber \\
\left[\left(\begin{array}{cc}
    E & 0 \\
    0 & E
\end{array}\right) - \left(\begin{array}{cc}
    \epsilon & 0 \\
    0 & \epsilon
\end{array}\right)\right]\left(\begin{array}{cc}
    C_{8,\uparrow\uparrow}  & 0 \\
    0 & C_{8,\uparrow\downarrow}
\end{array}\right)
= \left(\begin{array}{cc}
   t & 0 \\
    0 & t
\end{array}\right) \left(\begin{array}{cc}
   C_{7,\uparrow\uparrow}  & 0 \\
    0 & C_{7,\uparrow\downarrow}
\end{array}\right) + \left(\begin{array}{cc}
   t & 0 \\
    0 & t
\end{array}\right) \left(\begin{array}{cc}
   C_{9,\uparrow\uparrow}  & 0 \\
    0 & C_{9,\uparrow\downarrow}
\end{array}\right) + \left(\begin{array}{cc}
   t_D & 0 \\
    0 & t_D
\end{array}\right) \left(\begin{array}{cc}
        \tau_{\uparrow\uparrow}e^{ika} \\
    \tau_{\uparrow\downarrow}e^{ika}
    \end{array}\right)\nonumber \\
\left[\left(\begin{array}{cc}
    E & 0 \\
    0 & E
\end{array}\right) - \left(\begin{array}{cc}
    \epsilon & 0 \\
    0 & \epsilon
\end{array}\right)\right] \left(\begin{array}{cc}
   C_{9,\uparrow\uparrow}  & 0 \\
    0 & C_{9,\uparrow\downarrow}
\end{array}\right) = \left(\begin{array}{cc}
    t & 0 \\
    0 & t
\end{array}\right) \left(\begin{array}{cc}
   C_{8,\uparrow\uparrow}  & 0 \\
    0 & C_{8,\uparrow\downarrow}
\end{array}\right) + \left(\begin{array}{cc}
    t & 0 \\
    0 & t
\end{array}\right) \left(\begin{array}{cc}
   C_{10,\uparrow\uparrow}  & 0 \\
    0 & C_{10,\uparrow\downarrow}
\end{array}\right)\nonumber \\
\left[\left(\begin{array}{cc}    
    E & 0 \\
    0 & E
\end{array}\right) - \left(\begin{array}{cc}
    \epsilon & 0 \\
    0 & \epsilon
\end{array}\right)\right] \left(\begin{array}{cc}
   C_{10,\uparrow\uparrow}  & 0 \\
    0 & C_{10,\uparrow\downarrow}
\end{array}\right) = \left(\begin{array}{cc}
    t & 0 \\
    0 & t
\end{array}\right) \left(\begin{array}{cc}
   C_{11,\uparrow\uparrow}  & 0 \\
    0 & C_{11,\uparrow\downarrow}
\end{array}\right) + \left(\begin{array}{cc}
    t & 0 \\
    0 & t
\end{array}\right) \left(\begin{array}{cc}
   C_{13,\uparrow\uparrow}  & 0 \\
    0 & C_{13,\uparrow\downarrow}
\end{array}\right) + \left(\begin{array}{cc}
    t & 0 \\
    0 & t
\end{array}\right) \left(\begin{array}{cc}
   C_{9,\uparrow\uparrow}  & 0 \\
    0 & C_{9,\uparrow\downarrow}
\end{array}\right) \nonumber \\
\left[\left(\begin{array}{cc}
    E & 0 \\
    0 & E
\end{array}\right) - \left(\begin{array}{cc}
	\epsilon + h_{11}\cos\vartheta_{11} & \sin\vartheta_{11}e^{-i\varphi_{11}} \\ 
	\sin\vartheta_{11}e^{i\varphi_{11}} & \epsilon - h_{11}\cos\vartheta_{11}
\end{array}\right)\right] \left(\begin{array}{cc}
   C_{11,\uparrow\uparrow}  & 0 \\
    0 & C_{11,\uparrow\downarrow}
\end{array}\right)
  = \left(\begin{array}{cc}
    t & 0 \\
    0 & t
\end{array}\right) \left(\begin{array}{cc}
   C_{10,\uparrow\uparrow}  & 0 \\
    0 & C_{10,\uparrow\downarrow}
\end{array}\right) +  \left(\begin{array}{cc}
    t & 0 \\
    0 & t
\end{array}\right) \left(\begin{array}{cc}
   C_{12,\uparrow\uparrow}  & 0 \\
    0 & C_{12,\uparrow\downarrow}
\end{array}\right)\nonumber \\
\left[\left(\begin{array}{cc}    
    E & 0 \\
    0 & E
\end{array}\right) - \left(\begin{array}{cc}
    \epsilon & 0 \\
    0 & \epsilon
\end{array}\right)\right] \left(\begin{array}{cc}
    C_{12,\uparrow\uparrow}  & 0 \\
    0 & C_{12,\uparrow\downarrow}
\end{array}\right) = \left(\begin{array}{cc}
    t & 0 \\
    0 & t
\end{array}\right) \left(\begin{array}{cc}
   C_{11,\uparrow\uparrow}  & 0 \\
    0 & C_{11,\uparrow\downarrow}
\end{array}\right) + \left(\begin{array}{cc}
    t & 0 \\
    0 & t
\end{array}\right) \left(\begin{array}{cc}
   C_{14,\uparrow\uparrow}  & 0 \\
    0 & C_{14,\uparrow\downarrow}
\end{array}\right) + \left(\begin{array}{cc}
    t & 0 \\
    0 & t
\end{array}\right) \left(\begin{array}{cc}
   C_{13,\uparrow\uparrow}  & 0 \\
    0 & C_{13,\uparrow\downarrow}
\end{array}\right) \nonumber \\
\left[\left(\begin{array}{cc}
    E & 0 \\
    0 & E
\end{array}\right) - \left(\begin{array}{cc}
	\epsilon + h_{13}\cos\vartheta_{13} & \sin\vartheta_{13}e^{-i\varphi_{13}} \\ 
	\sin\vartheta_{13}e^{i\varphi_{13}} & \epsilon - h_{13}\cos\vartheta_{13}
\end{array}\right)\right] \left(\begin{array}{cc}
   C_{13,\uparrow\uparrow}  & 0 \\
    0 & C_{13,\uparrow\downarrow}
\end{array}\right)
  = \left(\begin{array}{cc}
    t & 0 \\
    0 & t
\end{array}\right) \left(\begin{array}{cc}
   C_{12,\uparrow\uparrow}  & 0 \\
    0 & C_{12,\uparrow\downarrow}
\end{array}\right) +  \left(\begin{array}{cc}
    t & 0 \\
    0 & t
\end{array}\right) \left(\begin{array}{cc}
   C_{10,\uparrow\uparrow}  & 0 \\
    0 & C_{10,\uparrow\downarrow}
\end{array}\right)\nonumber \\
\left[\left(\begin{array}{cc}
    E & 0 \\
    0 & E
\end{array}\right) - \left(\begin{array}{cc}
    \epsilon & 0 \\
    0 & \epsilon
\end{array}\right)\right] \left(\begin{array}{cc}
   C_{14,\uparrow\uparrow}  & 0 \\
    0 & C_{14,\uparrow\downarrow}
\end{array}\right) = \left(\begin{array}{cc}
    t & 0 \\
    0 & t
\end{array}\right) \left(\begin{array}{cc}
   C_{12,\uparrow\uparrow}  & 0 \\
    0 & C_{12,\uparrow\downarrow}
\end{array}\right) + \left(\begin{array}{cc}
    t & 0 \\
    0 & t
\end{array}\right) \left(\begin{array}{cc}
   C_{1,\uparrow\uparrow}  & 0 \\
    0 & C_{1,\uparrow\downarrow}
\end{array}\right)\nonumber \\
\left[\left(\begin{array}{cc}
    E & 0 \\
    0 & E
\end{array}\right) - \left(\begin{array}{cc}
    \epsilon & 0 \\
    0 & \epsilon_0
\end{array}\right)\right] \left(\begin{array}{cc}
    \tau_{\uparrow\uparrow}e^{ika} \\
    \tau_{\uparrow\downarrow}e^{ika}
    \end{array}\right)
 = \left(\begin{array}{cc}
   t_D & 0 \\
    0 & t_D
\end{array}\right) \left(\begin{array}{cc}
   C_{8,\uparrow\uparrow}  & 0 \\
    0 & C_{8,\uparrow\downarrow}
\end{array}\right) + \left(\begin{array}{cc}
    t_0 & 0 \\
    0 & t_0
\end{array}\right) \left(\begin{array}{cc}
    \tau_{\uparrow\uparrow}e^{2ika} \\
    \tau_{\uparrow\downarrow}e^{2ika}
    \end{array}\right) \nonumber \\
\label{eq3}
\end{eqnarray}}
\end{widetext}
The parameters $t_S$ and $t_D$ represent the coupling between source-to-channel and
channel-to-drain, respectively. $\rho$ and $\tau$ are to the reflection and transmission
probabilities, respectively. $\vartheta_i$ be the polar angle whereas $\varphi_i$
is the azimuthal angle. $k$ is wave vector and $a$ is the atomic length. By solving 
Eq.~\ref{eq3}, we get bond current densities for sites $i$ and $i+1$ of the channel as:
\begin{equation}
J_{i,i+1\sigma\sigma'}=\frac{(2e/\hbar)\mbox{Im}\left[t\,C_{i,\sigma\sigma'}^*
C_{i+1,\sigma\sigma'} \right]}{(2e/\hbar)(1/2)t_0\sin(ka)}
\label{eq4}
\end{equation}
In the above expressions, $\sigma$ is used for the incident spin, while $\sigma'$ represents
the transmitting spin.

Similarly, for the down spin incidence, we get a set of equations like
Eq.~\ref{eq3} and we evaluate $J_{i\rightarrow i+1\downarrow\downarrow}$ and 
$J_{i\rightarrow i+1\downarrow\uparrow}$. With all these components of circular bond current
densities now we define the net bond current density $J_{i,i+1}$ as $J_{i,i+1} =
J_{i,i+1\uparrow\uparrow} + J_{i,i+1\uparrow\downarrow} +
J_{i,i+1\downarrow\downarrow} + J_{i, i+1\downarrow\uparrow}$.

\end{document}